# Faster Functional Clustering via Gaussian Mixture Models

Hien D. Nguyen†, Geoffrey J. McLachlan, Jeremy F. P. Ullmann, and Andrew L. Janke*

February 13, 2017

**Abstract**

Functional data analysis (FDA) is an important modern paradigm for handling infinite-dimensional data. An important task in FDA is model-based clustering, which organizes functional populations into groups via subpopulation structures. The most common approach for model-based clustering of functional data is via mixtures of linear mixed-effects models. The mixture of linear mixed-effects models (MLMM) approach requires a computationally intensive algorithm for estimation. We provide a novel Gaussian mixture model (GMM) characterization of the model-based clustering problem. We demonstrate that this GMM-based characterization allows for improved computational speeds over the MLMM approach when applied via available functions in the **R** programming environment. Theoretical considerations for the GMM approach are discussed. An example application to a dataset based upon calcium imaging in the larval zebrafish brain is provided as a demonstration of the effectiveness of the simpler GMM approach.

*Hien Nguyen is with the Department of Mathematics and Statistics, La Trobe University, Bundoora, Melbourne, Australia 3083. Geoffrey McLachlan is with the School of Mathematics and Physics, The University of Queensland, St. Lucia, Brisbane, Australia 4075. Andrew Janke is with the Centre for Advanced Imaging, The University of Queensland, St. Lucia, Brisbane, Australia 4075. Jeremy Ullmann is with the Department of Neurology, Boston Children's Hospital and Harvard Medical School, Boston, USA 02115. †Corresponding Author: Hien Nguyen (Email: h.nguyen5@latrobe.edu.au).

## 1 Introduction

Functional data analysis (FDA) is the leading paradigm for modeling the statistical properties of data that are observed from infinite-dimensional functional objects. The seminal textbook, Ramsay & Silverman (2005), presents a comprehensive introduction to the foundations of FDA. Applications and demonstrations of FDA to real-world problems can be found in Ramsay & Silverman (2002). Nonparametric approaches to FDA are reviewed and discussed in Ferraty & Vieu (2006) and a user-friendly guide to conducting FDA with **R** is presented in Ramsay et al. (2009); see R Core Team (2016) regarding the **R** programming language.

When presented with data, a common task that arises is to organize the data into clusters, or groups of objects that are similar to one another (Jain & Dubes, 1988). The process of organizing data into clusters is known as clustering. Model-based clustering is the process of using a finite mixture of probability models to distinguish different subpopulation structures, which are declared as clusters; see McLachlan & Basford (1988). A recent survey of functional data clustering is presented in Jacques & Preda (2014a). A specific review of model-based clustering methods for functional data is provided below.

A mixture of linear mixed-effects models (MLMM) was applied by Celeux et al. (2005) to analyze time-course microarray data that are sampled at discrete time points. MLMMs with B-spline, Fourier, and wavelet bases were also considered by James & Sugar (2003), Ng et al. (2006), and Giacofci et al. (2013), respectively. Autoregressive extensions of the Ng et al. (2006) approach were considered by Scharl et al. (2010) and Wang et al. (2012), who were able to model both mean and variance effects. A two-dimensional extension of the James & Sugar (2003) methodology was considered by Nguyen et al. (2016b), for clustering random surfaces. A mixture-of-experts approach using linear segments was also considered by Same et al. (2011) for the clustering of electrical power signals.

The fitting of a MLMM requires a specialized and computationally-intensive EM (expectation–maximization) algorithm for maximum likelihood (ML) estimation; see McLachlan & Krishnan (2008) regarding EM algorithms. In the **R** programming language, there are a number of functions in specialized packages that can fit MLMMs. These functions include the **FLXMRlmm** driver within the **flexmix** package (Grun & Leisch, 2008), the **regmixEM.mixed** within the **mixtools** package (Benaglia et al., 2009), and the **emmixwire** function within the **EMMIXcontrasts** package (Ng et al., 2014, 2015).

We propose an alternative characterization to the MLMM approach via ordinary least squares (OLS) estimation of basis coefficients. This characterization converts the MLMM to the Gaussian mixture model (GMM), which can be fitted using a faster and more computationally-efficient EM algorithm for ML estimation. Such EM algorithms have been implemented in various **R** packages; for example **EMMIX** (McLachlan et al., 1999), **flexmix** (Grun & Leisch, 2008), **mixtools** (Benaglia et al., 2009), and **mclust** (Fraley & Raftery, 2003).

Theoretical results regarding the consistency and asymptotic normality of the ML estimator is presented. These theoretical results cover the case of correlated data, as well as the usual independent and identically distributed sampling scenario. An assessment of the efficacy of our new approach is conducted via a series of simulation studies. A real data analysis of a zebrafish calcium imaging data set is presented to demonstrate the methodology and its effectiveness.

As this article focuses only on the computational improvement that is possible via the conversion from the MLMM characterization to the GMM characterization, we will not be discussing alternative model-based approaches to FDA. Some interesting alternatives include the mixture of multivariate functional principal component analyses approach of Jacques & Preda (2014b) and the discriminative functional mixture approach of Bouveyron et al. (2015).

The paper proceeds as follows. The GMM approach to the MLMM is presented in Section 2. The ML estimation for the GMM is discussed in Section 3. The results from a series of simulation studies are presented in Section 4, and an analysis of an fMRI data set is demonstrated in Section 5. Conclusions are drawn in Section 6.

# 2 GMM Approach to the Characterization an MLMM

Let $\mathbb{Y}_n = \{Y_1(t), Y_2(t), ..., Y_n(t)\}$ be a sample of $n$ functions that are indexed by $t \in \mathbb{T}$, where $\mathbb{T}$ is an interval on $\mathbb{R}$. Suppose that we observe each function $Y_i(t)$ at $m$ indices $t_1, ..., t_m \in \mathbb{T}$, for $i = 1, ..., n$, indirectly via the corrupted variable

$$Z_i(t_j) = Y_i(t_j) + E_i(t_j), \tag{1}$$

where $E_i(t_j)$ is a random normal error with mean zero and some variance $\sigma^2 > 0$, and $j = 1, ..., m$.

In order to cluster $\mathbb{Y}_n$, we require an implicit model for the heterogeneity of the sample. This can be achieved via a linear parametric model of the form $Y_i(t) = \boldsymbol{B}_i^\top \boldsymbol{x}(t)$, where $\boldsymbol{B}_i \in \mathbb{R}^d$ is a random vector of appropriate dimensionality that captures the heterogeneity of the sample $\mathbb{Y}_n$. The $d$-dimensional function $\boldsymbol{x}(t)$ is a set of bases that are sufficiently-rich in the class of continuous functions. For example, we can take $\boldsymbol{x}(t)$ to be the first-$d$ monomials, a set of $d$ Fourier bases, or a $(d-2)$-knots B-spline system; see Ramsay & Silverman (2005, Sec. 3.3–3.6) for details and suggestions. The superscript $\top$ indicates matrix transposition.

The subpopulation heterogeneity of $\mathbb{Y}_n$ can now be imposed by supposing that the random vector $\boldsymbol{B}_i$ arises from a $g$-component GMM with density function

$$f(\boldsymbol{b};\boldsymbol{\theta}) = \sum_{c=1}^{g} \pi_c \phi(\boldsymbol{b};\boldsymbol{\mu}_c,\mathbf{V}_c), \tag{2}$$

where $\pi_c > 0$ and $\sum_{c=1}^{g} \pi_c = 1$, $\boldsymbol{\mu}_c \in \mathbb{R}^d$, and $\mathbf{V}_c \in \mathbb{R}^{d\times d}$ is a positive-definite matrix, for each $c = 1, ..., g$. Here, the $\pi_c$ are the prior component probabilities and

$$\phi(\boldsymbol{b};\boldsymbol{\mu}_c,\mathbf{V}_c) = |2\pi\mathbf{V}_c|^{-1/2} \exp\left[-\frac{1}{2}(\boldsymbol{b}-\boldsymbol{\mu}_c)^\top \mathbf{V}_c^{-1}(\boldsymbol{b}-\boldsymbol{\mu}_c)\right] \tag{3}$$

is the Gaussian density function with mean $\boldsymbol{\mu}_c$ and covariance matrix $\mathbf{V}_c$. The vector $\boldsymbol{\theta}$ stores the parameter components $\pi_c$, $\boldsymbol{\mu}_c$, and $\mathbf{V}_c$, for $c = 1, ..., g$.

For each $i = 1, ..., n$, we can write the relationship (1) across all $m$ indices as

$$\boldsymbol{Z}_i = \mathbf{X}\boldsymbol{B}_i + \boldsymbol{E}_i, \tag{4}$$

where $\boldsymbol{Z}_i^\top = (Z_i(t_1), ..., Z_i(t_m))$, $\boldsymbol{E}_i^\top = (E_i(t_1), ..., E_i(t_m))$, and

$$\mathbf{X}^\top = \left[\begin{array}{ccc} \boldsymbol{x}(t_1) & ... & \boldsymbol{x}(t_m) \end{array}\right].$$

From (4), we can write the OLS estimator of $\boldsymbol{B}_i$ as

$$\tilde{\boldsymbol{B}}_i = \left(\mathbf{X}^\top\mathbf{X}\right)^{-1}\mathbf{X}^\top\boldsymbol{Z}_i. \tag{5}$$

**Theorem 1.** *Under assumptions (1) and (2), $\tilde{\boldsymbol{B}}_i$ as defined by (5) has density*

*function*

$$f\left(\tilde{\boldsymbol{b}}_i;\boldsymbol{\vartheta}\right)=\sum_{c=1}^{k}\pi_c\phi_d\left(\tilde{\boldsymbol{b}}_i;\boldsymbol{\mu}_c,\mathbf{V}_c+\sigma^2\left(\mathbf{X}^\top\mathbf{X}\right)^{-1}\right), \tag{6}$$

*where we put the parameter components* $\pi_c$*,* $\boldsymbol{\mu}_c$*,* $\mathbf{V}_c$*, and* $\sigma^2$ *into* $\boldsymbol{\vartheta}$*.*

*Proof.* Start by expanding (5) to get $\tilde{\boldsymbol{B}}_i=\left(\mathbf{X}^\top\mathbf{X}\right)^{-1}\mathbf{X}^\top\left(\boldsymbol{Y}_i+\boldsymbol{E}_i\right)$, where $\boldsymbol{Y}_i^\top=\left(Y_i\left(t_1\right),...,Y_i\left(t_m\right)\right)$ and $\boldsymbol{E}_i^\top=\left(E_i\left(t_1\right),...,E_i\left(t_m\right)\right)$. Under the expansion $Y_i\left(t\right)=\boldsymbol{B}_i^\top\boldsymbol{x}\left(t\right)$, we obtain $\boldsymbol{Y}_i=\mathbf{X}\boldsymbol{B}_i$ and thus we can write

$$\begin{aligned}\tilde{\boldsymbol{B}}_i &= \left(\mathbf{X}^\top\mathbf{X}\right)^{-1}\mathbf{X}^\top\left(\mathbf{X}\boldsymbol{B}_i+\boldsymbol{E}_i\right)\\ &= \boldsymbol{B}_i+\left(\mathbf{X}^\top\mathbf{X}\right)^{-1}\mathbf{X}^\top\boldsymbol{E}_i.\end{aligned}$$

Since $E_i\left(t_j\right)$ is Gaussian with mean 0 and variance $\sigma^2$, we get the density of $\tilde{\boldsymbol{E}}=\left(\mathbf{X}^\top\mathbf{X}\right)^{-1}\mathbf{X}^\top\boldsymbol{E}_i$ to be Gaussian with mean $\mathbf{0}$ and covariance matrix $\left(\mathbf{X}^\top\mathbf{X}\right)^{-1}\mathbf{X}^\top\left(\sigma^2\mathbf{I}\right)\mathbf{X}\left(\mathbf{X}^\top\mathbf{X}\right)^{-1}=\sigma^2\left(\mathbf{X}^\top\mathbf{X}\right)^{-1}$, where $\mathbf{I}$ is an identity matrix of appropriate dimension. Since $\tilde{\boldsymbol{B}}_i=\boldsymbol{B}_i+\tilde{\boldsymbol{E}}_i$, we can write

$$f\left(\tilde{\boldsymbol{b}}_i;\boldsymbol{\vartheta}\right) = \sum_{c=1}^{k}\pi_c\int_{\mathbb{R}^d}\phi\left(\tilde{\boldsymbol{b}}_i+\tilde{\boldsymbol{e}};\boldsymbol{\mu}_c,\mathbf{V}_c\right)\phi\left(\tilde{\boldsymbol{e}};\mathbf{0},\sigma^2\left[\mathbf{X}^\top\mathbf{X}\right]^{-1}\right)\mathrm{d}\tilde{\boldsymbol{e}}.$$

Standard theory regarding the convolution of two Gaussian densities yields the desired result [e.g. see Vinga & Almeida (2004, Supp. Mat.)]. □

**Corollary 1.** *Adopt the same assumptions as Theorem 1. If* $\mathbf{V}_c+\sigma^2\left(\mathbf{X}^\top\mathbf{X}\right)^{-1}$ *is positive-definite, then we can write the density of* $\tilde{\boldsymbol{B}}_i$ *as*

$$f\left(\tilde{\boldsymbol{b}}_i;\boldsymbol{\psi}\right)=\sum_{c=1}^{k}\pi_c\phi\left(\tilde{\boldsymbol{b}}_i;\boldsymbol{\mu}_c,\boldsymbol{\Sigma}_c\right), \tag{7}$$

*where we put the parameter components* $\pi_c$*,* $\boldsymbol{\mu}_c$*, and* $\boldsymbol{\Sigma}_c=\mathbf{V}_c+\sigma^2\left(\mathbf{X}^\top\mathbf{X}\right)^{-1}$ *into* $\boldsymbol{\psi}$*.*

*Remark* 1. Corollary 1 implies that one can perform clustering under assumptions (1) and (2) in two steps. That is, in the first step, the OLS estimates can be computed for each $i = 1, ..., n$. And in the second step, the GMM (7) can be fitted and used for clustering.

*Remark* 2. In the sequel, we choose to adopt arbitrary component covariance matrices in (7) rather than to use the specific form of $\mathbf{\Sigma}_c$ as given by $\mathbf{V}_c + \sigma^2 \left(\mathbf{X}^\top \mathbf{X}\right)^{-1}$. As we care only about the estimation of a distribution for the OLS estimates $\tilde{\boldsymbol{B}}_i$ and not the functional observations $\boldsymbol{Y}_i$, the use of such a covariance structure has no effect on the performance of the GMM-based approach.

*Remark* 3. In the case when $\mathbf{X}^\top \mathbf{X}$ is singular, we can consider the Moore-Penrose inverse $\left(\mathbf{X}^\top \mathbf{X}\right)^{+}$ in place of $\left(\mathbf{X}^\top \mathbf{X}\right)^{-1}$, wherever it occurs (i.e (5), Theorem 1 and Corollary 1). As the Moore-Penrose inverse always exists, is unique, and is real whenever $\mathbf{X}^\top \mathbf{X}$ is real [cf. Seber (2008, Sec. 7.4)], its use in our application causes no issue.

*Remark* 4. Suppose that each function $Y_i(t)$ is sampled with independent noise at $m_i \in \mathbb{N}$ different points $\mathbb{T}_i = \{t_{i1}, ..., t_{im_i}\}$. Let $\boldsymbol{Z}_i^\top = (Z_i(t_1), ..., Z_i(t_{im_i}))$, $\boldsymbol{E}_i^\top = (E_i(t_1), ..., E_i(t_{im_i}))$, and

$$\mathbf{X}_i^\top = \begin{bmatrix} \boldsymbol{x}(t_{i1}) & ... & \boldsymbol{x}(t_{im_i}) \end{bmatrix},$$

where $Z_i(t)$ is as in (1), and like (4), we have $\boldsymbol{Z}_i = \mathbf{X}_i \boldsymbol{B}_i + \boldsymbol{E}_i$. If we naturally assume that the sampling of the time points $\mathbb{T}_i$, for each $i$, occurs with enough structure such that $\mathbf{X}_i^\top \mathbf{X}_i \to \mathbf{\Delta}$ in probability, as $n \to \infty$, for some invertible and positive-definite matrix $\mathbf{\Delta}$, then under construction (1) and (2), the OLS estimator $\tilde{\boldsymbol{B}}_i = \left(\mathbf{X}_i^\top \mathbf{X}_i\right)^{-1} \mathbf{X}_i^\top \boldsymbol{Z}_i$ converges in distribution to a mixture of form (6), where $\mathbf{X}^\top \mathbf{X}$ is replaced by $\mathbf{\Delta}$.

# 3 Maximum Likelihood Estimation of the GMM

Suppose that we observe $m$ noise corrupted realizations $z_i(t_1), ..., z_i(t_m)$ of the function evaluates $y_i(t_1), ..., y_i(t_m)$ $(i = 1, ...n)$, for each of the random functions in $\mathbb{Y}_n$. We can write $\boldsymbol{z}_i^\top = (z_i(t_1), ..., z_i(t_m))$ for each $i$.

As noted in Remark 1, we can implement the GMM-based approach (7) in two steps. On the first step, we compute the OLS estimates $\tilde{\boldsymbol{b}}_i = \left(\mathbf{X}^\top\mathbf{X}\right)^{-1}\mathbf{X}^\top\boldsymbol{z}_i$. On the second step, we fit the GMM (7) via ML estimation. That is, we seek an appropriate root first-order condition (FOC) $\nabla\ell_n(\boldsymbol{\psi}) = \mathbf{0}$, where

$$\ell_n(\boldsymbol{\psi}) = \sum_{i=1}^{n}\log f\left(\tilde{\boldsymbol{b}}_i;\boldsymbol{\psi}\right), \tag{8}$$

is the log-likelihood function given the data $\tilde{\boldsymbol{b}}_1$,..., $\tilde{\boldsymbol{b}}_n$, assuming that the functions in $\mathbb{Y}_n$ are independent.

*Remark* 5. If the functions in $\mathbb{Y}_n$ are not independent, then (8) can be viewed as the marginal log-likelihood function, and can be used for estimation under the maximum composite likelihood framework; see Lindsay (1988) and Varin (2008) regarding composite likelihood methods. An example use of maximum marginal log-likelihood estimation for clustering can be found in Nguyen et al. (2016a).

From Equations (3) and (7) we note that (8) has the log-sum-exponential form. This implies that a closed form solution to the FOC cannot be obtained [cf. Day (1969, Sec. 4)]. We can solve the FOC in an iterative manner via an EM algorithm instead.

## 3.1 EM Algorithm

Let $\boldsymbol{\psi}^{(0)}$ be some initial value and let $\boldsymbol{\psi}^{(r)}$ be the $r$th iterate of the EM algorithm, which contains the parameter components $\pi_c^{(r)}$, $\boldsymbol{\mu}_c^{(r)}$, and $\boldsymbol{\Sigma}_c^{(r)}$. On the $(r+1)$ th

E-step, for each $i$ and $c$, we compute the posterior probabilities

$$\tau_i^{(r+1)} = \pi_c^{(r)} \phi \left( \tilde{\boldsymbol{b}}_i; \boldsymbol{\mu}_c^{(r)}, \boldsymbol{\Sigma}_c^{(r)} \right) / f \left( \tilde{\boldsymbol{b}}_i; \boldsymbol{\psi}^{(r)} \right) = \mathbb{P}_{\boldsymbol{\psi}^{(r)}} \left( C_i = c | \tilde{\boldsymbol{B}}_i = \tilde{\boldsymbol{b}}_i \right),$$

where $C_i \in \{1, ..., g\}$ is the random variable that places $Y_i(t)$ into subpopulation $c$ with probability $\pi_c$.

On the $(r+1)$ th M-step, for each $c$, we compute the parameter updates

$$\pi_c^{(r+1)} = n^{-1} \sum_{i=1}^{n} \tau_i^{(r+1)},$$

$$\boldsymbol{\mu}_c^{(r+1)} = \left( \sum_{i=1}^{n} \tau_i^{(r+1)} \right)^{-1} \left( \sum_{i=1}^{n} \tau_i^{(r+1)} \tilde{\boldsymbol{b}}_i \right),$$

and

$$\boldsymbol{\Sigma}_c^{(r+1)} = \left( \sum_{i=1}^{n} \tau_i^{(r+1)} \right)^{-1} \left( \sum_{i=1}^{n} \tau_i^{(r+1)} \left[ \tilde{\boldsymbol{b}}_i - \boldsymbol{\mu}_c^{(r+1)} \right] \left[ \tilde{\boldsymbol{b}}_i - \boldsymbol{\mu}_c^{(r+1)} \right]^{\top} \right).$$

The E- and M-steps are alternated and iterated until some convergence criterion is reached; for example, a common convergence criterion is to stop when $\ell_n \left( \boldsymbol{\psi}^{(r+1)} \right) - \ell_n \left( \boldsymbol{\psi}^{(r)} \right) < \text{TOL}$, for some small $\text{TOL} > 0$. Upon convergence, the final iterate of the algorithm is declared as the ML estimate and is denoted by $\hat{\boldsymbol{\psi}}_n$, which contains the parameter components $\hat{\pi}_{c,n}$, $\hat{\boldsymbol{\mu}}_{c,n}$, and $\hat{\boldsymbol{\Sigma}}_{c,n}$.

*Remark* 6. As with all EM algorithms, the sequence $\ell_n \left( \boldsymbol{\psi}^{(r)} \right)$ is monotonically increasing in $r$. Further, let $\boldsymbol{\psi}^{(r)} \to \boldsymbol{\psi}^{(\infty)}$ as $r \to \infty$, or equivalently as $\text{TOL} \to 0$. If $\boldsymbol{\psi}^{(\infty)}$ is finite for some initial value $\boldsymbol{\psi}^{(0)}$, then $\boldsymbol{\psi}^{(\infty)}$ is a stationary point of (8). These results are obtained via validating the assumptions of Razaviyayn et al. (2013, Thm. 1).

*Remark* 7. The presented EM algorithm requires matrix inversion of $c$ $d \times d$-matrices in the E-step, and requires $c$ summations over $n$ outer-products of $d$-dimensional vectors in the M-step. If these matrix operations are bottlenecks due to large values of $d$, then a matrix operation-free algorithm can be used instead; see Nguyen & McLachlan (2015).

## 3.2 Large-Sample Theory

The large-sample theory of GMMs has been well-established; for example, see Redner & Walker (1984), van de Geer (1997), and Atienza et al. (2007). We present a useful extremum estimator theorem for clustering dependent data, which is based on Amemiya (1985, Thm. 4.1.2). The proof follows closely that of Nguyen & McLachlan (2015, Thm. 5) and Nguyen et al. (2016a, Thm. 2), and is thus omitted.

**Theorem 2.** *Let $\boldsymbol{Z}_1, ..., \boldsymbol{Z}_n$ be an identically distributed, stationary, and ergodic (or $\alpha$-mixing) sample of noisy observations of $\mathbb{Y}_n$, such that for each $i = 1, ..., n$, $\boldsymbol{Z}_i$ arises from a population that is characterized by (1) and (2). Let $\boldsymbol{\psi}_0$ be a strict-local maximizer of $\mathbb{E} \log f\left(\tilde{\boldsymbol{b}}; \boldsymbol{\psi}\right)$, where $f\left(\tilde{\boldsymbol{b}}; \boldsymbol{\psi}\right)$ is given by Equation (7). If $\Psi_n = \{\boldsymbol{\psi} : \nabla \ell_n = \mathbf{0}\}$ (where we take $\Psi_n = \left\{\bar{\boldsymbol{\psi}}\right\}$, for some valid $\bar{\boldsymbol{\psi}}$, if $\nabla \ell_n = \mathbf{0}$ has no solution), then for any $\epsilon > 0$,*

$$\lim_{n \to \infty} \mathbb{P}\left[\inf_{\boldsymbol{\psi} \in \Psi_n} \left(\boldsymbol{\psi} - \boldsymbol{\psi}_0\right)^T \left(\boldsymbol{\psi} - \boldsymbol{\psi}_0\right) > \epsilon\right] = 0.$$

*Remark* 8. As remarked in Amemiya (1985, Sec. 4.1), Theorem 2 only states that there exists a consistent root to the FOC equation. It is well known that the log-likelihoods for GMMs are both multimodal and unbounded. As such, Theorem 2 is a useful result and guarantees that some solution of the FOC equation will be convergent to a parameter vector that resembles the generative model. Due to the multimodality of the log-likelihood, multiple-restart techniques are required to search the parameter space; some useful techniques are described in McLachlan & Peel (2000, Sec. 2.12.2).

*Remark* 9. Define an $M$-dependent sample $\mathbb{Y}_n$ as one where for all $i, i' = 1, ..., n$, $Y_i(t)$ and $Y_{i'}(t)$ are independent if $|i - i'| > M$. An $M$-dependent sample is an $\alpha$-mixing sample and thus satisfies the hypothesis of Theorem 2 [cf. Bradley (2005, Sec. 2.1)]. An assumption of $M$-dependence is generally admissible for image analysis tasks and is thus useful for such applications.

### 3.3 Clustering Rule

Following the approach of McLachlan & Basford (1988), we say that

$$\hat{c}_i = \arg\max_{c=1,\ldots,g} \hat{\tau}_{i.n} \tag{9}$$

is the cluster allocation of $Y_i(t)$ $(i = 1, ..., n)$, where

$$\hat{\tau}_{i,n} = \hat{\pi}_{c,n}\phi\left(\tilde{\boldsymbol{b}}_i; \hat{\boldsymbol{\mu}}_{c,n}, \hat{\boldsymbol{\Sigma}}_{c,n}\right)/f\left(\tilde{\boldsymbol{b}}_i; \hat{\boldsymbol{\psi}}_n\right) = \mathbb{P}_{\hat{\boldsymbol{\psi}}_n}\left(C_i = c|\tilde{\boldsymbol{B}}_i = \tilde{\boldsymbol{b}}_i\right)$$

is the posterior probability of $Y_i(t)$ belonging to class $c$, based on the data $\boldsymbol{Z}_i$ and the ML estimator $\hat{\boldsymbol{\psi}}_n$. Via continuous mapping, if $\hat{\boldsymbol{\psi}}_n$ is a consistent estimator of $\boldsymbol{\psi}_0$, then $\hat{c}_{i,n} \to c_i$ as $n \to \infty$, where $c_i$ is the Bayes' optimal allocation of $Y_i(t)$ [cf. McLachlan (1992, Sec. 1.4)].

## 4 Simulation Studies

We perform a pair of simulation studies (S1 and S2) to compare the performances of the MLMM and GMM approaches on some datasets. We use **R** functions to fit the MLMMs, as per James & Sugar (2003), an an EM algorithm to fit the GMM characterization that we have presented in Sections 2 and 3.

### 4.1 S1

Three functions for clustering via the MLMM-based approach and three functions for clustering via the GMM-based approach were assessed. The EM algorithm for the fitting of the MLMM was implemented via the **emmixwire** function (within **EMMIXcontrasts**), the **regmixEM.mixed** function (**mixtools**), and the **stepFlexmix** function with the **FLXMRlmm** driver (**flexmix**). The EM algorithm for the GMM was implemented via the **Mclust** function (**mclust**), the **mvnormalmixEM** function (**mixtools**), and the **stepFlexmix** function with the **FLXMCmvnorm** driver (**flexmix**). All settings were left as default unless necessary to enforce assumptions (1) and (2). OLS estimations that were required for the GMM approach were performed using the **smooth.basisPar** function from the **R** package: **fda** (Ramsay et al., 2009). Where possible, the stopping criterion was set with TOL = 0.01. Since all settings were left default, the choice of a routine for searching among local maxima was left to each of the function libraries. For example, the **Mclust** utilizes a hierarchical preclustering, whereas **stepFlexmix** simply randomizes initial assignments.

We simulated $n = 30, 60, 150, 300$ noise-corrupted functions $Z_i(t)$ $(i = 1, ..., n)$ at $m = 10, 20, 50, 100$ indices in $\mathbb{T} = [-1, 1]$. Of the $n$ functions, equal numbers are simulated with mean models $\mathbb{E}_c[Y(t)] = \boldsymbol{\mu}_c^\top \boldsymbol{x}(t)$, for $c = 1, 2, 3$, respectively, where $\boldsymbol{\mu}_1^\top = (0, 0, 0, 0, 0)$, $\boldsymbol{\mu}_2^\top = (0, 1, 0, 0, 0)$, $\boldsymbol{\mu}_3^\top = (0, -1, 0, 0, 0)$, and $\boldsymbol{x}^\top(t) = \left(1, t, t^2, t^3, t^4\right)$. The random-effects covariances and noise variance were set to $\mathbf{V}_1 = \mathbf{V}_2 = \mathbf{V}_3 = \text{diag}\left(0.05^2\right)$, and $\sigma^2 = 0.1^2$, respectively. Each combination of $m$ and $n$ were replicated 100 times, and each of the clustering functions were run on each replication with the number of components set to $g = 3$. An example of the $m = 50$ and $n = 30$ case is visualized in Figure 1.

Each of the seven clustering functions were assessed on two properties: clustering accuracy and computational speed. Clustering accuracy is assessed via the adjusted-Rand index (ARI) of Hubert & Arabie (1985). The ARI measures the agreement of the obtained clustering from each function versus the known generative subpopulation that each simulated function arises from, in that it measures the number of times that pairs of observations are in agreement. The ARI takes values between -1 and 1, with 1 indicating a perfectly-matched clustering.

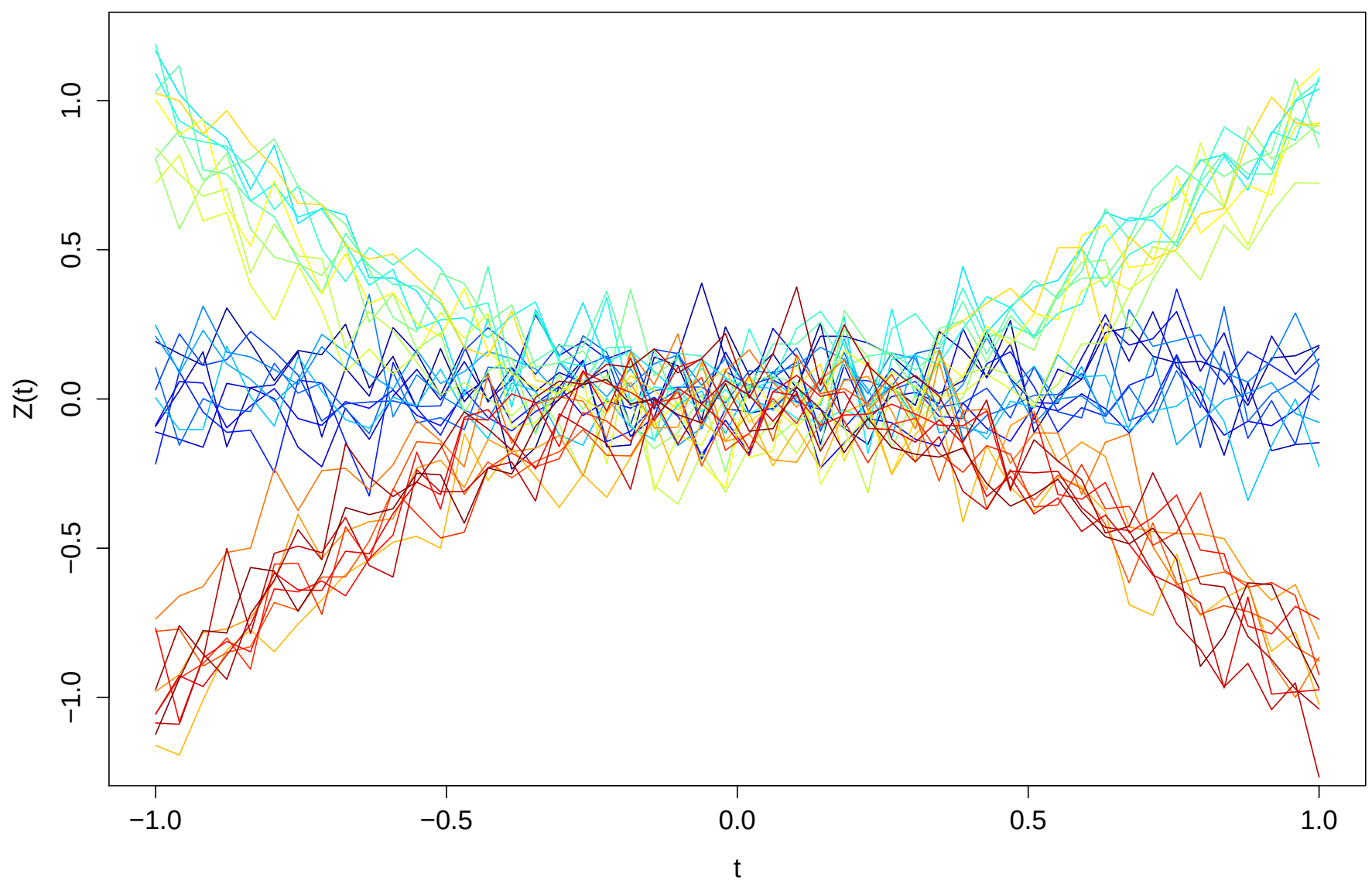

Figure 1: Noise-corrupted functions from a replication of the $m = 50$ and $n = 30$ case in S1. Each color indicates a different function.

The computational speed was measured using the **proc.time** function in **R**. The OLS estimation times were included in the timing of the GMM-based functions. All computations were conducted within the **R** programming environment [version 3.0.2; see R Core Team (2016)] on a Windows PC with an Intel Core i7-2600 CPU running at 3.40 GHz with 16 GB internal RAM.

The results of this study are presented in Tables 1 and 2. From Table 1, it can be observed that **emmixwire** is the fastest and the most accurate function for conducting MLMM-based clustering, over all scenarios. It can be over 450 times faster (**stepFlexmix**; $m = 10$, $n = 300$) and at least seven times faster (**stepFlexmix**; $m = 100$, $n = 30$), than the next-best MLMM-based function. It can also be observed that **emmixwire** achieves a perfect average ARI values (1.00) across all scenarios, with the next-best method achieving at best an average ARI of 0.93 (**regmixEM.mixed**; $m = 10$, $n = 300$).

From Table 2, it can be observed that for the implementation of the GMM-based clustering the **Mclust** function was the most accurate, with perfect average ARI values for all but five cases, where the worst average ARI value was 0.74. The next-best average ARI value was 0.66 (**stepFlexmix**; $m = 20$, $n = 300$). The **Mclust** function was also the fastest in implementing the the GMM-based clustering.

We conclude from the results in Tables 1 and 2 that the **emmixwire** and **Mclust** are the best performing MLMM-based and GMM-based functions, re-spectively. Comparing the approaches, we observed that **emmixwire** was more accurate than **Mclust** in five scenarios, whereas the **Mclust** was at least five times faster than **emmixwire**, on average ($m = 10$, $n = 150$).

Table 1: Average time and ARI results of MLMM-based clustering functions for S1.

| | | **emmixwire** | | **regmixEM.mixed** | | **stepFlexmix** | |
|---|---|---|---|---|---|---|---|
| $m$ | $n$ | Time (s) | ARI | Time (s) | ARI | Time (s) | ARI |
| 10 | 30 | 0.07 | 1.00 | 10.18 | 0.66 | 5.81 | 0.01 |
| 10 | 50 | 0.08 | 1.00 | 25.81 | 0.80 | 11.30 | 0.02 |
| 10 | 150 | 0.11 | 1.00 | 81.77 | 0.88 | 25.36 | 0.13 |
| 10 | 300 | 0.17 | 1.00 | 186.65 | 0.93 | 78.00 | 0.19 |
| 20 | 30 | 0.09 | 1.00 | 11.74 | 0.54 | 4.27 | 0.02 |
| 20 | 50 | 0.10 | 1.00 | 28.97 | 0.71 | 7.72 | 0.04 |
| 20 | 150 | 0.16 | 1.00 | 90.36 | 0.85 | 23.67 | 0.11 |
| 20 | 300 | 0.27 | 1.00 | 225.84 | 0.91 | 78.28 | 0.10 |
| 50 | 30 | 0.19 | 1.00 | 19.76 | 0.45 | 3.90 | 0.03 |
| 50 | 50 | 0.23 | 1.00 | 56.02 | 0.62 | 7.02 | 0.09 |
| 50 | 150 | 0.37 | 1.00 | 176.23 | 0.80 | 27.86 | 0.10 |
| 50 | 300 | 0.70 | 1.00 | 404.27 | 0.85 | 93.42 | 0.09 |
| 100 | 30 | 0.66 | 1.00 | 69.96 | 0.47 | 5.17 | 0.06 |
| 100 | 50 | 0.74 | 1.00 | 196.19 | 0.49 | 10.48 | 0.08 |
| 100 | 150 | 1.06 | 1.00 | 610.98 | 0.72 | 41.29 | 0.06 |
| 100 | 300 | 1.55 | 1.00 | 1253.83 | 0.83 | 124.65 | 0.06 |

## 4.2 S2

From S1, we have found **emmixwire** and **Mclust** to be the best performing MLMM-based and GMM-based functions, respectively. In order to assess which is faster and more accurate, we constructed the following simulation scenarios.

We simulated $n = 250, 500, 1000, 2500$ noise-corrupted functions $Z_i(t)$ ($i = 1, ..., n$) at $m = 50, 100, 200, 500$ indices in $\mathbb{T} = [0, 2\pi]$. Of the $n$ functions, equal numbers were simulated with mean models $\mathbb{E}_c[Y(t)] = \boldsymbol{\mu}_c^\top \boldsymbol{x}(t)$, for $c = 1, ..., 5$, respectively, where $\boldsymbol{\mu}_1^\top = (0, 0, ..., 0)$, $\boldsymbol{\mu}_2^\top = (1, 0, ..., 0)$, $\boldsymbol{\mu}_3^\top = (-1, 0, ..., 0)$,

Table 2: Average time and ARI results of GMM-based clustering functions for S1.

| | | **Mclust** | | **mvnormalmixEM** | | **stepFlexmix** | |
|---|---|---|---|---|---|---|---|
| $m$ | $n$ | Time (s) | ARI | Time (s) | ARI | Time (s) | ARI |
| 10 | 30 | 0.00 | 0.74 | 0.70 | 0.24 | 0.20 | 0.07 |
| 10 | 50 | 0.01 | 0.84 | 2.48 | 0.19 | 0.38 | 0.11 |
| 10 | 150 | 0.02 | 1.00 | 10.59 | 0.37 | 0.86 | 0.32 |
| 10 | 300 | 0.06 | 1.00 | 24.10 | 0.53 | 1.26 | 0.64 |
| 20 | 30 | 0.01 | 0.91 | 0.66 | 0.17 | 0.21 | 0.07 |
| 20 | 50 | 0.01 | 0.97 | 2.26 | 0.19 | 0.38 | 0.11 |
| 20 | 150 | 0.02 | 1.00 | 10.63 | 0.36 | 0.84 | 0.32 |
| 20 | 300 | 0.06 | 1.00 | 25.33 | 0.50 | 1.28 | 0.66 |
| 50 | 30 | 0.01 | 0.99 | 0.74 | 0.16 | 0.18 | 0.06 |
| 50 | 50 | 0.01 | 1.00 | 2.09 | 0.13 | 0.36 | 0.12 |
| 50 | 150 | 0.02 | 1.00 | 8.66 | 0.41 | 0.76 | 0.38 |
| 50 | 300 | 0.07 | 1.00 | 20.66 | 0.42 | 1.14 | 0.58 |
| 100 | 30 | 0.01 | 1.00 | 0.71 | 0.14 | 0.19 | 0.08 |
| 100 | 50 | 0.01 | 1.00 | 2.18 | 0.13 | 0.35 | 0.10 |
| 100 | 150 | 0.04 | 1.00 | 9.31 | 0.32 | 0.83 | 0.40 |
| 100 | 300 | 0.06 | 1.00 | 22.63 | 0.46 | 1.12 | 0.66 |

$\boldsymbol{\mu}_4^\top = (0, 1, ..., 0)$, $\boldsymbol{\mu}_5^\top = (0, -1, ..., 0)$, an

$$\boldsymbol{x}^\top (t) = (1, \sin t, \cos t, \sin 2t, \cos 2t, \sin 3t, \cos 3t, \sin 4t, \cos 4t) .$$

The random-effects covariances and noise variance were set to $\mathbf{V}_1 = ... = \mathbf{V}_5 = \text{diag}\left(0.25^2\right)$, and $\sigma^2 = 0.5^2$, respectively. Each combination of $m$ and $n$ were replicated 100 times, and each of the clustering functions were run on each replication with the number of components set to $g = 5$. An example of the $m = 50$ and $n = 250$ case is visualized in Figure 2.

The computation setting is exactly the same as in S1. The average computation times and ARI values are reported in Table 3.

From Table 3, we observe that the GMM approach as implemented by the **Mclust** function is comparable to the MLMM approach implemented using the **emmixwire** ($m = 50$, $n = 2500$) and can be over 700 times faster ($m = 500$, $N = 250$). Furthermore, we note that the computation speed for the GMM approach appears to depend very little on $m$ due to the speed of conducting OLS estimation. In terms of accuracy, the two methods are comparable for the majority of scenarios, with differences in average ARI values that are mostly less than 0.10. However, the GMM approach obtains average ARI values that are greater than 0.30 more than the MLMM approach in two cases ($m = 500$, $n = 250$, and $m = 500$, $n = 500$). This indicates that the GMM-based approach was both faster and more accurate across the tested scenarios of S2, than the MLMM-based approach.

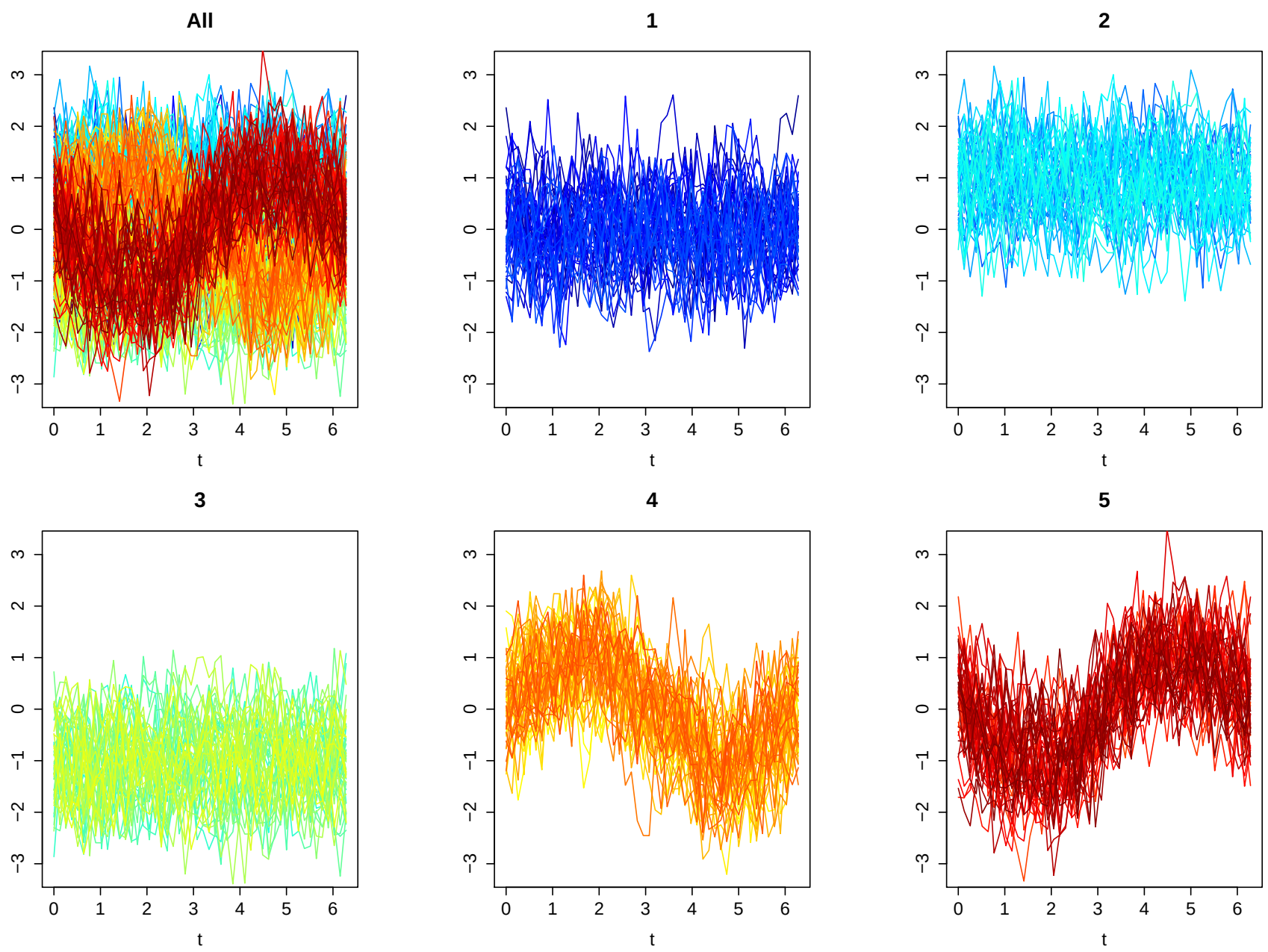

Figure 2: Noise-corrupted functions from a replication of the $m = 50$ and $n = 250$ case in S2. All functions are plotted in the first panel, with each color indicating a different function. In the numbered panels, only the functions with respective mean models are plotted.

Table 3: Average time and ARI results of clustering functions for S2.

| | | **emmixwire** | | **mclust** | |
|---|---|---|---|---|---|
| $m$ | $n$ | Time (s) | ARI | Time (s) | ARI |
| 50 | 250 | 0.78 | 0.80 | 0.09 | 0.75 |
| 50 | 500 | 1.33 | 0.82 | 0.27 | 0.78 |
| 50 | 1000 | 2.97 | 0.85 | 0.99 | 0.84 |
| 50 | 2500 | 10.70 | 0.87 | 10.02 | 0.87 |
| 100 | 250 | 1.85 | 0.79 | 0.09 | 0.76 |
| 100 | 500 | 3.08 | 0.82 | 0.25 | 0.82 |
| 100 | 1000 | 6.47 | 0.86 | 0.99 | 0.86 |
| 100 | 2500 | 24.37 | 0.88 | 10.00 | 0.88 |
| 200 | 250 | 7.17 | 0.69 | 0.09 | 0.78 |
| 200 | 500 | 10.35 | 0.80 | 0.24 | 0.84 |
| 200 | 1000 | 19.39 | 0.85 | 0.99 | 0.87 |
| 200 | 2500 | 58.22 | 0.89 | 10.03 | 0.89 |
| 500 | 250 | 72.84 | 0.38 | 0.10 | 0.77 |
| 500 | 500 | 83.32 | 0.47 | 0.27 | 0.84 |
| 500 | 1000 | 109.20 | 0.70 | 1.02 | 0.88 |
| 500 | 2500 | 220.13 | 0.83 | 10.09 | 0.90 |

# 5 Example Application

To demonstrate the use of the GMM characterization for clustering functional data, we consider an analysis of a time series data set arising from the calcium imaging of a zebrafish brain. The calcium imaging was performed on a 5 day post fertilization *Tg(HuC:GCamp5)* (Akerboom et al., 2012) zebrafish brain; see Muto & Kawakami (2013) for an example, regarding the calcium imaging of zebrafish brains. These data had been previously considered in Nguyen et al. (2016a).

The images were acquired on an inverted spinning disk microscope, more specifically, an Axio Observer Z1 (Carl Zeiss) equipped with a CSU-W1 spinning-disk head (Yokogawa Corporation of America), ORCA-Flash4.0 v2 sCMOS camera (Hamamatsu Photonics), 20x 0.8 NA PlanApo and 40x 1.2 NA C-Apo objectives. Image acquisition was performed using SlideBook 6.0 (3I, Inc). The zebrafish was subjected to pharmacologically induced neuronal activation over 500 seconds, and a single plane of $1024 \times 1024$ time-series data were acquired at 10 Hz over this time period. The data were down-sampled to a $512 \times 512$ image in space and the time series were smoothed and down-sampled to length $m = 500$ in time. The time domain was normalized to $\mathbb{T} = [0, 1]$. Using a numerical threshold, the image was manually masked to produce a final set of $n = 23445$ time series from pixels displaying interesting neuronal activity. Mean and standard deviation images (over time) of the data are presented in Figures 3 and 4, respectively.

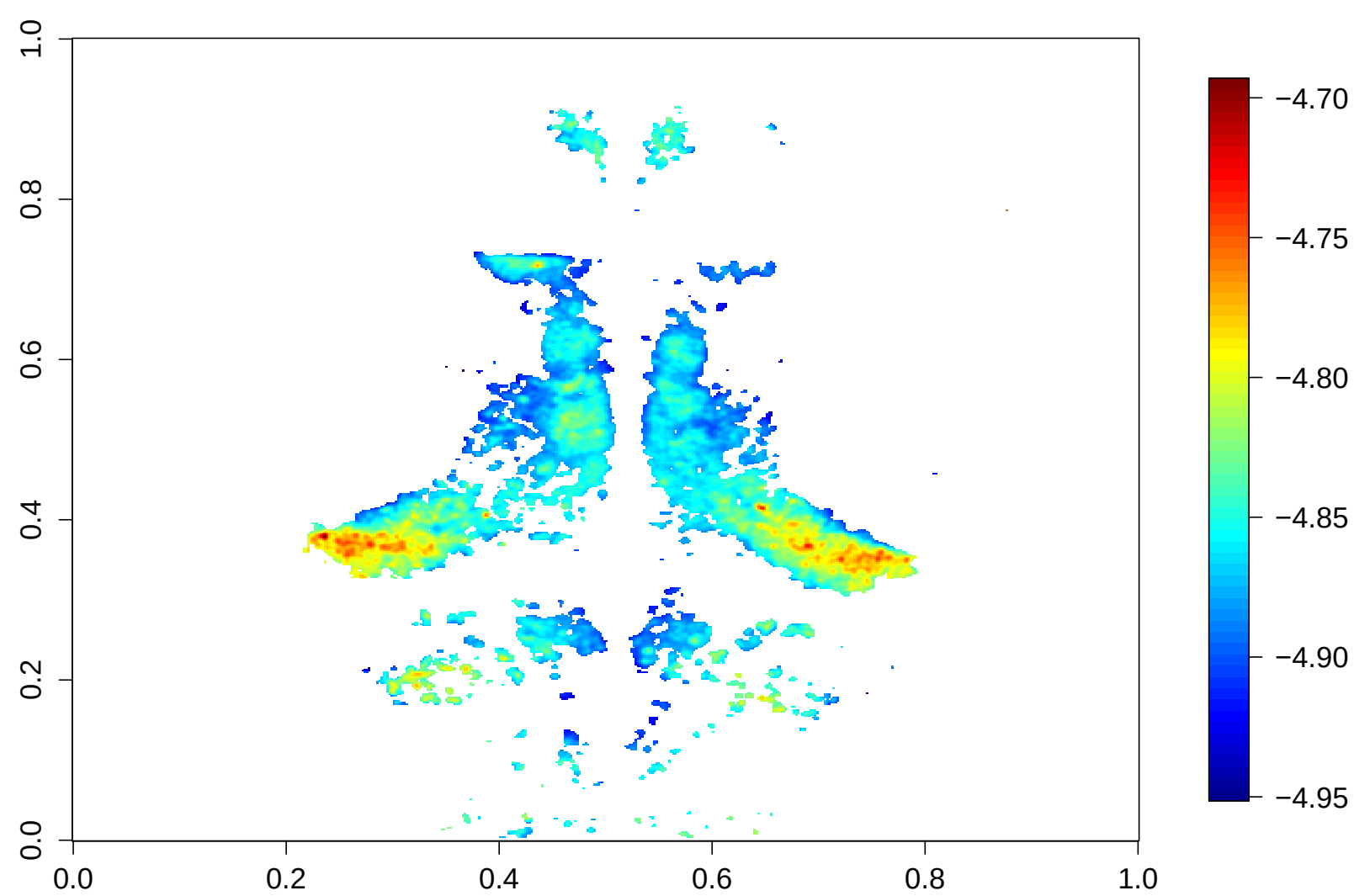

Figure 3: Mean plot of the manually masked zebrafish calcium image. The color of the pixel indicates the mean over the 500 time periods at each pixel.

To cluster the pixels of the image, we treated the time series at each pixel as a function and utilized a $(d-2)$-node B-spline system ($d = 20$) to model each function. Upon obtaining the OLS estimates for each pixel, we estimated $g = 1, ..., 60$ component GMMs to cluster the functions at each pixel.

Using the data-driven slope estimation (DDSE) technique of Baudry et al. (2010), we obtained the model selection criterion

$$\text{Criterion}\,(g) = -n^{-1}\ell_n\left(\boldsymbol{\psi}_n^{[g]}\right) + 3.55 \times 10^{-4} \times \text{Pen}\,(g)\,, \tag{10}$$

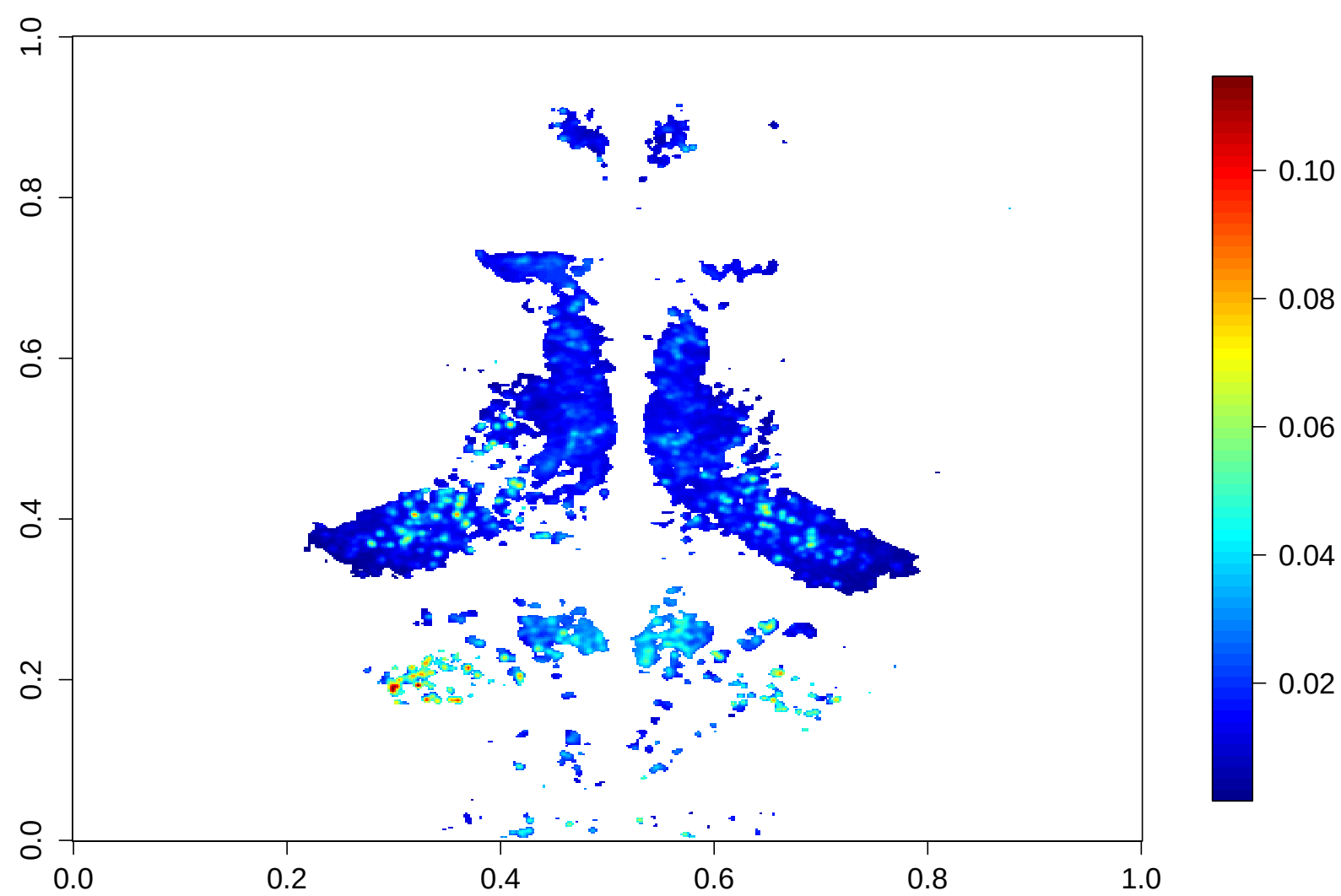

Figure 4: Standard deviation plot of the manually masked zebrafish calcium image. The color of the pixel indicates the standard deviation over the 500 time periods at each pixel.

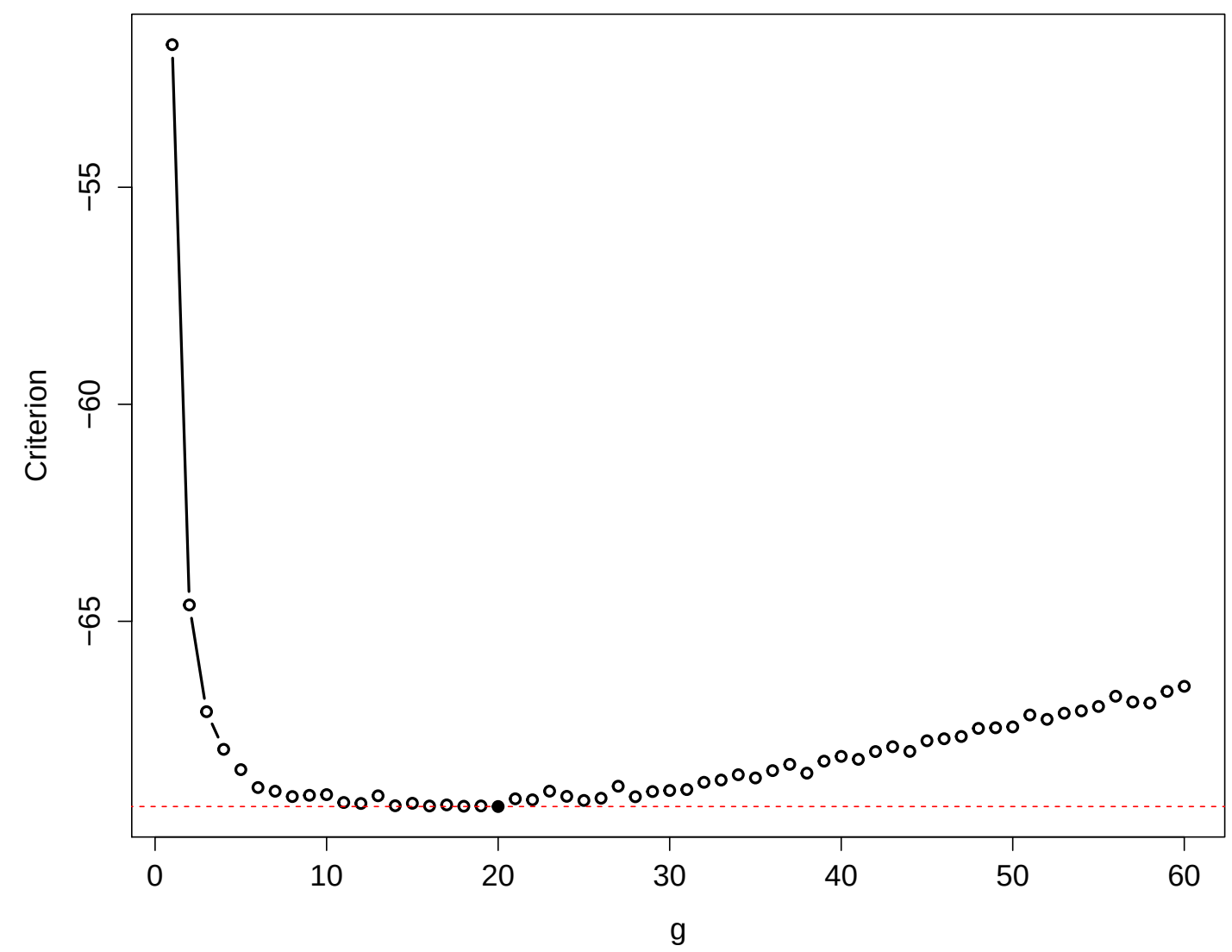

Figure 5: The value of the criterion (10), graphed against the number of components $g$ of the fitted GMMs. The solid dot indicates the optimal model. The dashed line indicates the optimal criterion value.

where $\text{Pen}\,(g) = g\,[1 + d + (d+1)\,d/2] - 1$ is the number of parameter components and $\boldsymbol{\psi}_n^{[g]}$ is the ML estimate of a $g$-component GMM. The form of the penalty is as suggested in Baudry et al. (2010, Table 1); see also Maugis & Michel (2011). Like the Bayesian information criterion of Schwarz (1978), the best model is the one that minimizes the criterion. In the case of the analyzed data, the optimal model was $g = 20$. Figure 5 graphs the criterion versus the number of components of the fitted GMM.

The clustering into $g = 20$ clusters via rule (9) is displayed in Figure 6. The clusters are ordered by the average of the estimated mean functions $\hat{\mathbb{E}}_c\,[Y\,(t)] = \hat{\boldsymbol{\mu}}_{c,n}^{\top}\boldsymbol{x}\,(t)$, for $c = 1, ..., 20$. The estimated mean functions for each of the 20 clusters are displayed in Figure 7.

From Figure 6, we notice that the cluster allocations appear to be spatially coherent, which is an indication that they are identifying biologically relevant features in the images. The arrangements of the spotted arrangements of the allocations are consistent with the variability patterns of Figure 4 and the progression of the cluster allocations are consistent with the means in Figure 3. Figure 7 illustrates that the clustering allocations organize the pixels into similarly shaped functional groups. The $g = 20$ clusters found are consistent with the 17 clusters found in Nguyen et al. (2016a). Differences between the clusterings are due to the fact that Nguyen et al. (2016a) cluster the time series in the frequency domain, whereas this example performs clustering in the time domain.

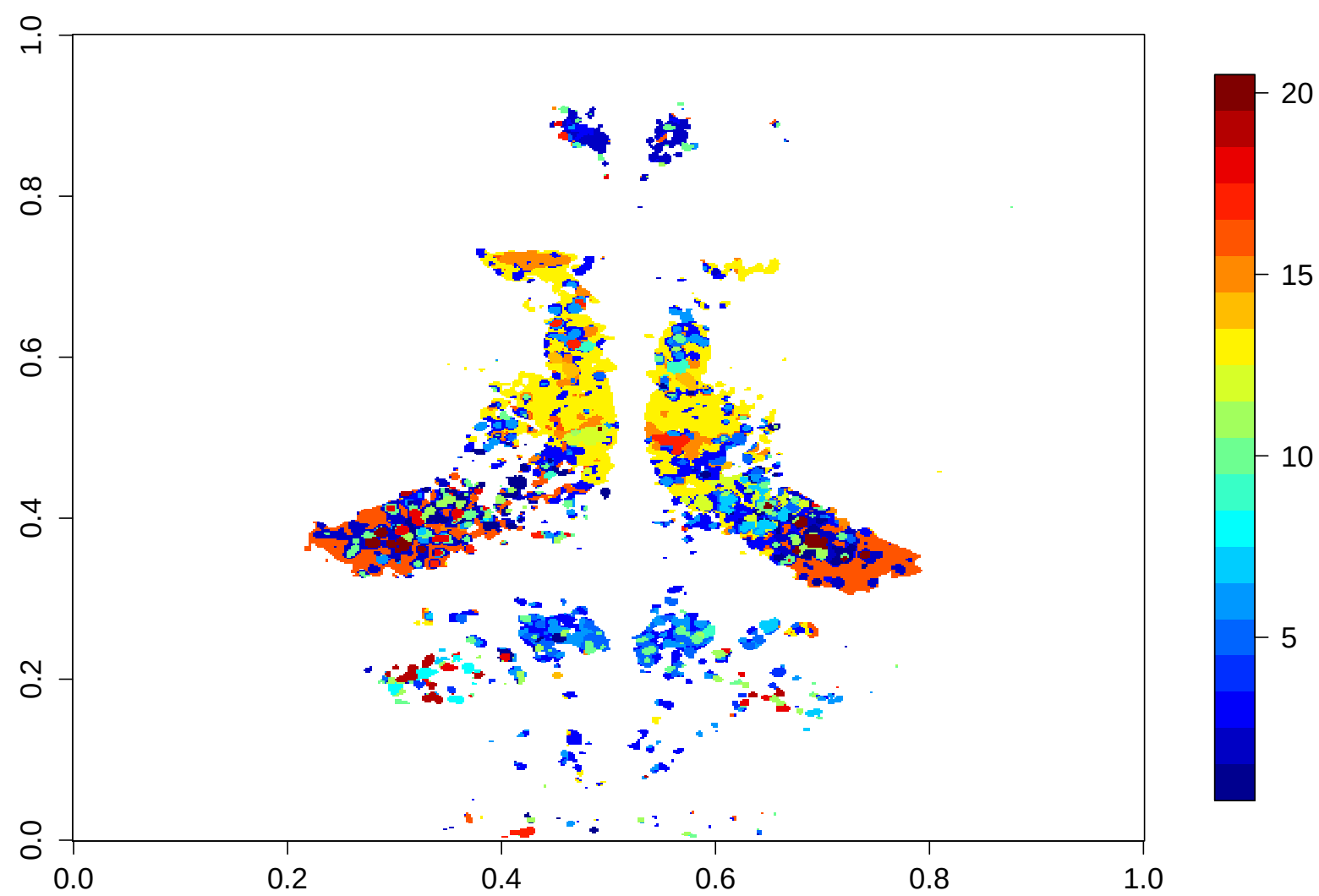

Figure 6: Clustering of the manually masked zebrafish calcium image. The color of the pixel indicates its allocation with respect to to one of the $g = 20$ clusters.

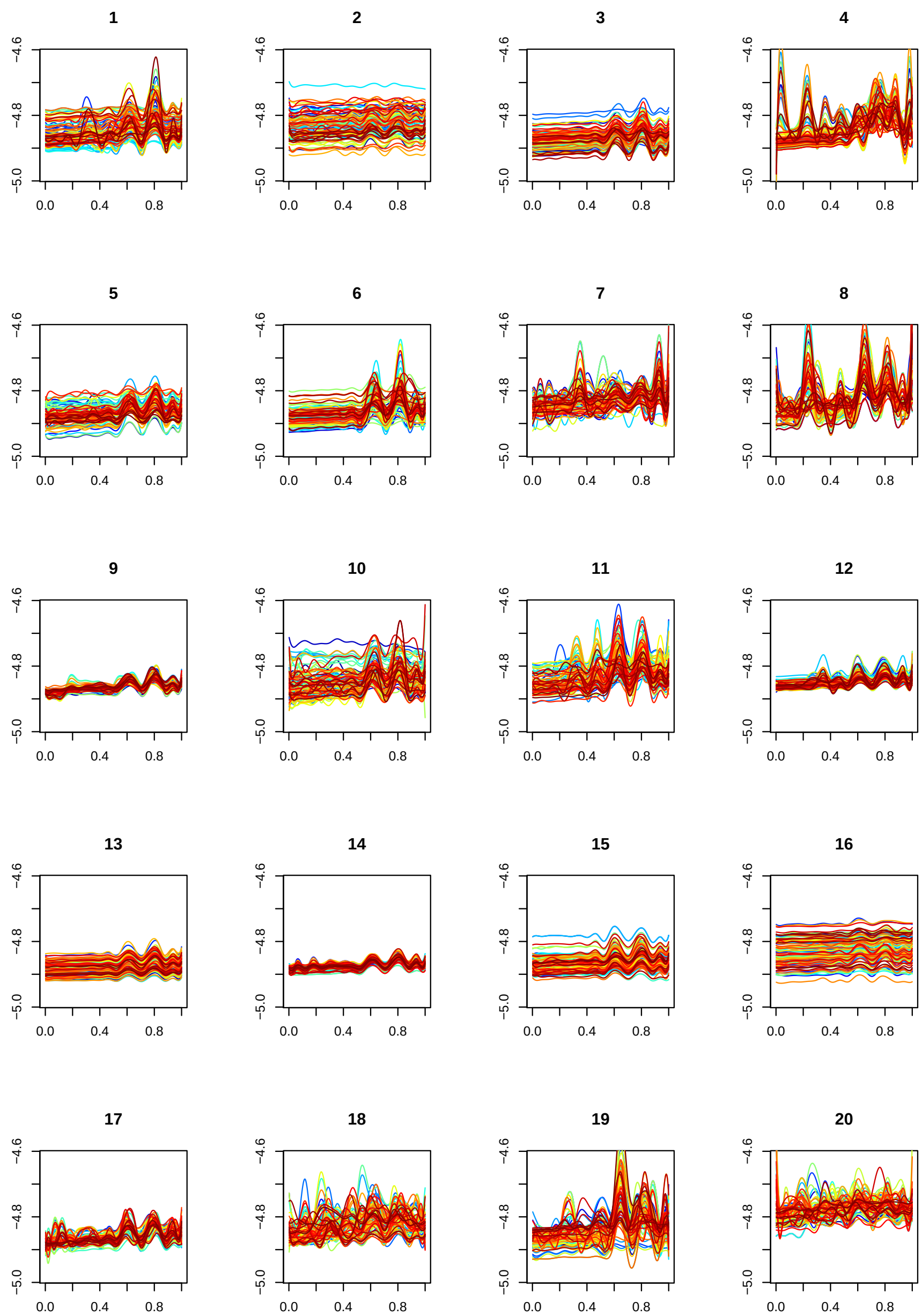

Figure 7: Estimated mean functions for each of the $g = 20$ clusters are displayed in black. The colored curves are a random sample of 100 OLS B-spline estimates of functions belonging to each cluster. The abscissa and ordinate of each subgraph indicates the time and image signals, respectively. Subgraphs are labeled as per Figure 6.

# 6 Conclusions

FDA is an important modern paradigm for handling infinite-dimensional data. Model-based clustering techniques aim to provide a distributional framework for organizing functional data into groups.

The main approach to model-based clustering of functional data is via MLMMs. The estimation of MLMMs require a computationally intensive EM algorithm. We demonstrate that the MLMM approach can be replaced via a GMM characterization instead. This allows for a more computationally efficient approach to model-based clustering of functional data.

In some simulations, we demonstrated that the GMM approach can provide a faster model-based clustering of functional data than the MLMM approach. Also, the GMM approach can provide a more effective clustering, particularly if the same size $n$ is not large relative to the number of indices $m$. The MLMM and GLM approaches can be implemented in the **R** programming language using the **emmixwire** and **Mclust** functions, respectively. A zebrafish calcium imaging example demonstrates the effectiveness of the GMM-based clustering approach in its application to a real dataset.

# References

Akerboom, J., Chen, T.-W., Wardill, T. J., Tian, L., Marvin, J. S., Mutlu, S., Calderon, N. C., Esposti, F., Borghuis, B. G., Sun, X. R., Gordus, A., Orger, M. B., Portugues, R., Engert, F., Macklin, J. J., Filosa, A., Aggarwal, A., Kerr, R. A., Takagi, R., Kracun, S., Shigetomi, E., Khakh, B. S., Baier, H., Lagnado, L., Wang, S. S.-H., Bargmann, C. I., Kimmel, B. E., Jayaraman, V., Svoboda, K., Kim, D. S., Schreiter, E. R., & Looger, L. L. (2012). Optimization of a GCaMP calcium indicator for neural activity imaging. *Journal of Neuroscience*, 32, 13819–13840.

Amemiya, T. (1985). *Advanced Econometrics*. Cambridge: Harvard University Press.

Atienza, N., Garcia-Heras, J., Munoz-Pichardo, J. M., & Villa, R. (2007). On the consistency of MLE in finite mixture models of exponential families. *Journal of Statistical Planning and Inference*, 137, 496–505.

Baudry, J.-P., Maugis, C., & Michel, B. (2010). *Slope heuristic: overview and implementation*. Technical Report hal-00461639, INRIA.

Benaglia, T., Chauveau, D., Hunter, D. R., & Young, D. S. (2009). mixtools: An R package for analyzing finite mixture models. *Journal of Statistical Software*, 32, 1–29.

Bouveyron, C., Come, E., & Jacques, J. (2015). The discriminative functional mixture model for a comparative analysis of bike sharing systems. *Annals of Applied Statistics*, 9, 1726–1760.

Bradley, R. C. (2005). Basic properties of strong mixing conditions. A survey and some open questions. *Probability Surveys*, 2, 107–144.

Celeux, G., Martin, O., & Lavergne, C. (2005). Mixture of linear mixed models for clustering gene expression profiles from repeated microarray experiments. *Statistical Modelling*, 5, 243–267.

Day, N. E. (1969). Estimating the components of a mixture of normal distributions. *Biometrika*, 56, 463–474.

Ferraty, F. & Vieu, P. (2006). *Nonparametric Functional Data Analysis*. New York: Springer.

Fraley, C. & Raftery, A. E. (2003). Enhanced model-based clustering, density estimation, and discriminant analysis software: MCLUST. *Journal of Classification*, 20, 263–286.

Giacofci, M., Lambert-Lacroix, S., Marot, G., & Picard, F. (2013). Wavelet-based clustering for mixed-effects functional models in high dimension. *Biometrics*, 69, 31–40.

Grun, B. & Leisch, F. (2008). Flexmix version 2: finite mixtures with concomitant variables and varying and constant parameters. *Journal of Statistical Software*, 28, 1–35.

Hubert, L. & Arabie, P. (1985). Comparing partitions. *Journal of Classification*, 2, 193–218.

Jacques, J. & Preda, C. (2014a). Functional data clustering: a survey. *Advances in Data Analysis and Classification*, 8, 231–255.

Jacques, J. & Preda, C. (2014b). Model-based clusterinf for multivariate functional data. *Computational Statistics and Data Analysis*, 71, 92–106.

Jain, J. K. & Dubes, R. C. (1988). *Algorithms for Clustering Data*. Englewood Cliffs: Prentice Hall.

James, G. M. & Sugar, C. A. (2003). Clustering for sparsely sampled functional data. *Journal of the American Statistical Association*, 98, 397–408.

Lindsay, B. (1988). Composite likelihood methods. *Contemporary Mathematics*, 8, 221–239.

Maugis, C. & Michel, B. (2011). A non asymptotic penalized criterion for Gaussian model selection. *ESAIM: Probability and Statistics*, 15, 41–68.

McLachlan, G. J. (1992). *Discriminant Analysis And Statistical Pattern Recognition*. New York: Wiley.

McLachlan, G. J. & Basford, K. E. (1988). *Mixture Models: Inference And Applications To Clustering*. New York: Marcel Dekker.

McLachlan, G. J. & Krishnan, T. (2008). *The EM Algorithm And Extensions*. New York: Wiley, 2 edition.

McLachlan, G. J. & Peel, D. (2000). *Finite Mixture Models*. New York: Wiley.

McLachlan, G. J., Peel, D., Basford, K. E., & Adams, P. (1999). The EMMIX software for the fitting of mixtures of normal and t-components. *Journal of Statistical Software*, 4(2).

Muto, A. & Kawakami, K. (2013). Prey capture in zebrafish larvae serves as a model to study cognitive functions. *Frontiers in Neural Circuits*, 7, 1–5.

Ng, S.-K., McLachlan, G., & Wang, K. (2014). *EMMIXcontrasts: Contrasts in mixed effects for EMMIX model with random effects*.

Ng, S. K., McLachlan, G. J., Ben-Tovim, K. W. L., & Ng, S. W. (2006). A mixture model with random-effects components for clustering correlated gene-expression profiles. *Bioinformatics*, 22, 1745–1752.

Ng, S.-K., McLachlan, G. J., Wang, K., Nagymanyoki, Z., Liu, S., & Ng, S.-W. (2015). Inference on differences between clclass using cluster-specific contrasts of mixed effects. *Biostatistics*, 16, 98–112.

Nguyen, H. D. & McLachlan, G. J. (2015). Maximum likelihood estimation of Gaussian mixture models without matrix operations. *Advances in Data Analysis and Classification*, 9, 371–394.

Nguyen, H. D., McLachlan, G. J., Ullmann, J. F. P., & Janke, A. L. (2016a). Spatial clustering of time-series via mixture of autoregressions models and Markov Random Fields. *Statistica Neerlandica*, In Press.

Nguyen, H. D., McLachlan, G. J., & Wood, I. A. (2016b). Mixture of spatial spline regressions for clustering and classification. *Computational Statistics and Data Analysis*, 93, 76–85.

R Core Team (2016). *R: a language and environment for statistical computing*. R Foundation for Statistical Computing.

Ramsay, J. O., Hooker, G., & Graves, S. (2009). *Functional Data Analysis with R and MATLAB*. New York: Springer.

Ramsay, J. O. & Silverman, B. W. (2002). *Applied Functional Data Analysis: Methods and Case Studies*. New York: Springer.

Ramsay, J. O. & Silverman, B. W. (2005). *Functional Data Analysis*. New York: Springer.

Razaviyayn, M., Hong, M., & Luo, Z.-Q. (2013). A unified convergence analysis of block successive minimization methods for nonsmooth optimization. *SIAM Journal of Optimization*, 23, 1126–1153.

Redner, R. A. & Walker, H. F. (1984). Mixture densities, maximum likelihood and the EM algorithm. *SIAM Review*, 26, 195–239.

Same, A., Chamroukhi, F., Govaert, G., & Aknin, P. (2011). Model-based clustering and segmentation of time series with change in regime. *Advances in Data Analysis and Classification*, 5, 301–321.

Scharl, T., Grun, B., & Leisch, F. (2010). Mixtures of regression models for time course gene expression data: evalluation of initialization and random effects. *Bioinformatics*, 26, 370–377.

Schwarz, G. (1978). Estimating the dimensions of a model. *Annals of Statistics*, 6, 461–464.

Seber, G. A. F. (2008). *A Matrix Handbook For Statisticians*. New York: Wiley.

van de Geer, S. (1997). Asymptotic normality in mixture models. *ESAIM: Probability and Statistics*, 1, 17–33.

Varin, C. (2008). On composite marginal likelihoods. *Advances in Statistical Analysis*, 92, 1–28.

Vinga, S. & Almeida, J. S. (2004). Renyi continuous entropy of DNA sequences. *Journal of Theoretical Biology*.

Wang, K., Ng, S. K., & McLachlan, G. J. (2012). Clustering of time-course gene expression profiles using normal mixed models with autoregressive random effects. *BMC Bioinformatics*, 13, 300.